\documentclass[pra,aps,amsmath,amssymb,eqsecnum,showkeys]{revtex4}
\newcommand{\hh}{{\mathcal{H}}}
\newcommand{\hk}{{\mathcal{K}}}
\newcommand{\lnp}{{\mathcal{L}}}
\newcommand{\lsp}{{\mathcal{L}}_{+}}
\newcommand{\id}{{\mathrm{id}}}

\newcommand{\sch}{{\mathrm{Sch}}}
\newcommand{\ron}{{\mathrm{ran}}}
\newcommand{\spc}{{\mathrm{spec}}}
\newcommand{\llv}{\Lambda}
\newcommand{\rrv}{\Upsilon}
\newcommand{\veps}{\varepsilon}
\newcommand{\pen}{\openone}
\newcommand{\niz}{{\mathbf{0}}}
\newcommand{\hta}{{\mathsf{A}}}
\newcommand{\htb}{{\mathsf{B}}}
\newcommand{\htx}{{\mathsf{X}}}
\newcommand{\hty}{{\mathsf{Y}}}
\newcommand{\htp}{{\mathsf{P}}}
\newcommand{\htq}{{\mathsf{Q}}}
\newcommand{\htk}{{\mathsf{K}}}
\newcommand{\htr}{{\boldsymbol{\rho}}}
\newcommand{\htv}{{\boldsymbol{\varrho}}}
\newcommand{\hts}{{\boldsymbol{\sigma}}}
\newcommand{\hto}{{\boldsymbol{\omega}}}
\newcommand{\htop}{{\boldsymbol{\varpi}}}
\newcommand{\htpi}{{\boldsymbol{\pi}}}
\newcommand{\htlam}{{\boldsymbol{\lambda}}}
\newcommand{\tr}{\mathrm{tr}}
\newcommand{\wb}{\widetilde{B}}

\begin{document}
\clearpage
\preprint{}

\title{On quantum conditional entropies defined in terms of the $f$-divergences}

\author{Alexey E. Rastegin}
\affiliation{Department of Theoretical Physics, Irkutsk State University,
Gagarin Bv. 20, Irkutsk 664003, Russia}

\begin{abstract}
We consider a family of quantum conditional entropies based on the
concept of quantum $f$-divergences. First, we explicitly formulate
conditions under which the notion of quantum conditional entropy
is well defined in this way. In particular, we demand that the
value of conditional entropy be independent of any extension of
the principal Hilbert space. Using fundamental properties of the
quantum $f$-divergences, several interesting relations for such
conditional entropies are formulated. We separately examine an
especially important case of quantum conditional entropies related
to the Tsallis divergence.
\end{abstract}


\maketitle

\pagenumbering{arabic}
\setcounter{page}{1}

\section{Introduction}

The concept of relative entropy is of great
importance in information theory. Many principal results in
quantum information theory are closely related to properties of
the relative entropy. Its monotonicity and joint
convexity are very essential \cite{vedral02,petz08}. The Shannon
entropy of probability distributions and the von Neumann entropy
of density matrices are commonly used measures of an informational
content. Other entropic functions have found to be useful in
various questions \cite{bengtsson}. There exist reasons to study
possible extensions of the standard divergence. Many of them
can be unified within the concept of $f$-divergences \cite{hmp10}.
This approach is a quantum counterpart of Csisz\'{a}r's
$f$-divergences \cite{ics67}. Petz's quasi-entropies are a primary 
example \cite{petz86}. Quantum divergences may be
adopted as distance measures, though they are not a metric in the
formal sense. In studying problems of information theory, a
collection of distinguishability measures is typically used
\cite{audn13}. For example, the authors of \cite{cbr13} recently
proposed one-shot generalizations for usual mutual information
related to the von Neumann entropy.

The conditional entropy is widely used in information theory
\cite{CT91}. In the classical regime, the conditional R\'{e}nyi
\cite{EP04} and Tsallis \cite{sf06} entropies are examples of
generalized conditional entropies. The R\'{e}nyi entropy
\cite{renyi61} and the Tsallis entropy \cite{tsallis} are well
known extensions of the Shannon entropy. Relations between
generalized conditional entropies and the error probability are of
interest in communication \cite{rastkyb}. In the quantum regime,
the notion of conditional entropy is also very important
\cite{levitin99,schrader00}. There are more than one way to fit an
approach to quantum conditional entropies. The quantum conditional
entropy is usually written as the difference of corresponding von
Neumann entropies \cite{lieb75}. It is closely related to the
strong subadditivity \cite{LR73}. The conditional entropy can also
be expressed in terms of the standard relative entropy
\cite{petz08,nielsen}. An approach based on relevant divergences
was developed in details for some interesting cases
\cite{rren05,krs09,mdsft13}. To study entanglement in a bipartite
system, the authors of \cite{giro13} proposed a modification of
the conditional entropy after a local measurement on one of the
subsystems.

The aim of the present paper is to study quantum conditional
entropies based on the notion of $f$-divergences. The paper is
organized as follows. In Sect. \ref{sec2}, the preliminary
material is given. In particular, we discuss required properties
of the quantum $f$-divergences. Corresponding conditional
entropies are defined in Sect. \ref{sec3}. We consider general
conditions, when the presented extension is well defined.
Particularly, we demand that the value of conditional entropy be
independent of any extension of the state space. When the function
used in the $f$-divergence enjoys certain conditions, the
considered entropic quantities obey some interesting relations.
These results are generally discussed in Sect. \ref{sec4}. In
Sect. \ref{sec5}, we separately analyze the case of Tsallis
entropies. Some additional properties of the conditional
$\alpha$-entropies of Tsallis type are derived. In Sect.
\ref{sec6}, we conclude the paper with a summary of results.

\section{Preliminaries}\label{sec2}

In this paper, we will study quantum conditional entropies defined
on base of the quantum $f$-divergences. The latter is a quantum
counterpart of the Csisz\'{a}r $f$-divergence \cite{ics67}. This
concept provide a unified approach to studying relative entropies
in the classical regime. Let $\xi\mapsto{f}(\xi)$ be a convex function
on $\xi\in[0;+\infty)$ with $f(1)=0$. The Csisz\'{a}r
$f$-divergence of the probability distribution $\{p_{x}\}$ from
$\{q_{x}\}$ is defined as \cite{ics67}
\begin{equation}
S_{f}(p{||}q):=\sum\nolimits_{x}{q_{x}{\,}{f}{\left(\frac{p_{x}}{q_{x}}\right)}}
{\,}. \label{cfdf}
\end{equation}
Using the Jensen inequality for a convex function, we easily
obtain
\begin{equation}
\sum\nolimits_{x}{q_{x}{\,}{f}{\left(\frac{p_{x}}{q_{x}}\right)}}\geq{f}{\left(\sum\nolimits_{x}{p_{x}}\right)}=f(1)
{\,}, \label{cfin}
\end{equation}
or merely $S_{f}(p{||}q)\geq0$. In other words, the $f$-divergence
takes nonnegative values, and $S_{f}(p{||}p)=0$. Several
approaches to defining divergences and other entropic measures are
reviewed in \cite{ics08}. Without explicit assignment of the
function, various inequalities with the Csisz\'{a}r
$f$-divergences were obtained in \cite{drag03}.

Let $\hh$ be a finite-dimensional Hilbert space. By $\lnp(\hh)$
and $\lsp(\hh)$, we denote the space of linear operators on $\hh$
and the set of positive semi-definite ones. Eigenvalues of an operator
$\htx\in\lnp(\hh)$ form its spectrum $\spc(\htx)$. By
$\ron(\htx)$, we mean the range of operator $\htx$. For positive
$\hta$, the subspace $\ron(\hta)$ is spanned by those eigenvectors
of $\hta$ that correspond to strictly positive eigenvalues. For
$\htx,\hty\in\lnp(\hh)$, we define the Hilbert--Schmidt inner
product
\begin{equation}
\langle\htx{\,},\hty\rangle_{\mathrm{hs}}:=\tr(\htx^{\dagger}\hty)
\ . \label{hsinp}
\end{equation}
In the following, we use the convention that powers of a positive
operator are taken only on its support. For any
$\hta\in\lsp(\hh)$, by $\hta^{0}$ we mean the orthogonal projector
onto $\ron(\hta)$. In the finite-dimensional case, we
further define $\htp\vee\htq$ as the projector onto the sum of
subspaces $\ron(\htp)+\ron(\htq)$. The last definition should be
modified in the infinite-dimensional case, since the sum of two
closed subspaces is not necessarily closed. In the following, we
restrict a consideration to finite dimensions.

Any state of a quantum system is described by density operator
$\htr\in\lsp(\hh)$. We usually deal with normalized states such
that $\tr(\htr)=1$. By $\htv_{*}=d^{-1}\pen$ with the identity
operator $\pen$, we mean the completely mixed state in $d$
dimensions. For density operators $\htr$ and $\hts$, the quantum
relative entropy is expressed as \cite{petz08}
\begin{equation}
D_{1}(\htr||\hts):=
\left\{
\begin{array}{ll}
\tr(\htr\ln\htr-\htr\ln\hts) {\>},& {\mathrm{if{\ }}}\ron(\htr)\subseteq\ron(\hts)
{\>}, \\
+\infty{\>}, & {\mathrm{otherwise}}{\>}.
\end{array}
\right.
\label{relan}
\end{equation}
There exist several generalizations of the quantity (\ref{relan}).
Many of them can be described as partial cases on the quantum
$f$-divergence \cite{hmp10}. Taking $\hta,\htb\in\lsp(\hh)$, we
introduce the left multiplication $\llv_{\hta}$ and the right
multiplication $\rrv_{\htb}$, namely
\begin{equation}
\llv_{\hta}:{\>} \htx\mapsto{\hta\htx}
\ , \qquad
\rrv_{\htb}:{\>} \htx\mapsto{\htx\htb}
\ , \label{lrmults}
\end{equation}
where $\htx\in\lnp(\hh)$. Left and right multiplications commute
with each other, i.e.,
$\llv_{\hta}{\,}\rrv_{\htb}=\rrv_{\htb}{\,}\llv_{\hta}$ for
$\hta,\htb\in\lsp(\hh)$. In terms of the corresponding projectors,
we write the spectral decompositions
\begin{equation}
\hta=\sum_{a\in\spc(\hta)}{a{\,}\htp_{a}}
\ , \qquad
\htb=\sum_{b\in\spc(\htb)}{b{\,}\htq_{b}}
\ . \label{spdab}
\end{equation}
Let $\xi\mapsto{f}(\xi)$ be a real-valued function on
$\xi\in[0;+\infty)$ such that it is continuous on $(0,+\infty)$
and the limit
\begin{equation}
\ell(f):=\underset{\xi\to{+\infty}}{\lim}{\>}\xi^{-1}f(\xi)
\label{fxlm}
\end{equation}
exists in $[-\infty;+\infty]$. Using the set
$\left\{ab^{-1}:{\>}a\in\spc(\hta),{\>}b\in\spc(\htb)\right\}$, we
write \cite{hmp10}
\begin{equation}
f(\llv_{\hta}{\,}\rrv_{\htb^{-1}}):=\sum_{a\in\spc(\hta)}{\,}\sum_{b\in\spc(\htb)}
{f(ab^{-1}){\,}\llv_{\htp_{a}}\rrv_{\htq_{b}}}
\ . \label{fabpq}
\end{equation}
In the right-hand side of (\ref{fabpq}), we assume that
$\ron(\hta)\subseteq\ron(\htb)$. Then the $f$-divergence of $A$
with respect to $B$ is defined by \cite{hmp10}
\begin{equation}
D_{f}(\hta||\htb):=
\left\langle\htb^{1/2},f(\llv_{\hta}{\,}\rrv_{\htb^{-1}}){\,}\htb^{1/2}\right\rangle_{\mathrm{hs}}
\ . \label{qsfdef}
\end{equation}
When $\ron(\hta)\nsubseteq\ron(\htb)$, the quantum $f$-divergence
is defined by the formula \cite{hmp10}
\begin{equation}
D_{f}(\hta||\htb):=\underset{\veps\searrow{0}}{\lim}{\>}D_{f}(\hta||\htb+\veps\pen)
\ . \label{qsfdep}
\end{equation}
Taking $f_{1}(\xi)=\xi\ln\xi$, we obtain the standard relative
entropy (\ref{relan}) rewritten for positive operators of any
trace. Namely, for $\ron(\hta)\subseteq\ron(\htb)$ we have
\begin{equation}
D_{1}(\hta||\htb):=
\tr(\hta\ln\hta-\hta\ln\htb)
\ . \label{qendf1}
\end{equation}
Basic properties of the $f$-divergence (\ref{qsfdep}) are examined
in \cite{hmp10}. Under the above conditions on the function
$\xi\mapsto{f}(\xi)$, the $f$-divergence is continuous in its
second entry \cite{hmp10}. It must be stressed that continuity in
the first entry does not hold in general \cite{hmp10}.

Similarly to (\ref{cfin}), we can check positivity of the
$f$-divergence in some important cases. Consider two normalized
states
\begin{equation}
\htr=\sum_{a\in\spc(\htr)}{a{\,}\htpi_{a}}
\ , \qquad
\hts=\sum_{b\in\spc(\hts)}{b{\,}\htlam_{b}}
\ . \label{rspt}
\end{equation}
In the case $\ron(\htr)\subseteq\ron(\hts)$, the divergence
(\ref{qsfdef}) can be represented as
\begin{equation}
D_{f}(\htr||\hts)=\sum_{a\in\spc(\htr)}
{\,}\sum_{b\in\spc(\hts)\setminus\{0\}}
{b{\,}f{\left(\frac{a}{b}\right)}
{\,}\tr(\htpi_{a}\htlam_{b})}
\ . \label{rspt1}
\end{equation}
This expression directly follows from item 2.3 of \cite{hmp10}.
The factors $b\tr(\htpi_{a}\htlam_{b})$ in the sum
(\ref{rspt1}) can be treated as probabilities due to the relation
\begin{equation}
\sum_{a\in\spc(\htr)}
{\,}\sum_{b\in\spc(\hts)\setminus\{0\}}{b{\>}\tr(\htpi_{a}\htlam_{b})}=\tr(\hts)=1
\ . \label{rspt2}
\end{equation}
Because of $\ron(\htr)\subseteq\ron(\hts)$, we further have
\begin{equation}
\sum_{a\in\spc(\htr)}
{\,}\sum_{b\in\spc(\hts)\setminus\{0\}}{a{\>}\tr(\htpi_{a}\htlam_{b})}=\tr(\htr)=1
\ . \label{rspt3}
\end{equation}
Combining these facts with (\ref{rspt1}) and convexity of the
function $\xi\mapsto{f}(\xi)$ finally gives
\begin{equation}
D_{f}(\htr||\hts)\geq{f}(1)
\ , \label{rspt4}
\end{equation}
whence $D_{f}(\htr||\hts)\geq0$ in the case $f(1)\geq0$. The
result (\ref{rspt4}) is a quantum counterpart of (\ref{cfin}). The
$f$-divergence of normalized $\htr$ with respect to normalized
$\hts$ is nonnegative, when $\ron(\htr)\subseteq\ron(\hts)$ and
the used convex function obeys $f(1)\geq0$. This conclusion can be
extended to any two operators $\hta,\htb\in\lsp(\hh)$ such that
$\ron(\hta)\subseteq\ron(\htb)$ and $\tr(\hta)=\tr(\htb)$. Here,
we merely use the homogeneity \cite{hmp10}
\begin{equation}
D_{f}(\lambda{\,}\hta||\lambda{\,}\htb)=\lambda{\,}D_{f}(\hta||\htb)
\ , \label{fhom}
\end{equation}
where $\lambda\in[0;+\infty)$. Approaching quantum conditional
entropies on base of the $f$-divergences, we will use operators
with unequal traces. In this case, corresponding divergences may
take negative values.

Some existing extensions of the relative entropy (\ref{relan})
do not directly follow from the definition (\ref{qsfdef}). The
R\'{e}nyi divergence is most important of them. In some respects,
the traditional form of quantum R\'{e}nyi's divergence is not
satisfactory. Recently, M\"{u}ller-Lennert et al. \cite{mdsft13}
and Wilde et al. \cite{wwy13} proposed a new definition of quantum
R\'{e}nyi's divergence. Further extensions of this approach are
considered in \cite{adat13}. The writers of \cite{wwy13} used the
``sandwiched'' relative R\'{e}nyi entropy in studying a strong
converse for the classical capacity of entanglement-breaking and
Hadamard channels. For the pure-loss bosonic channel, this issue
is examined in \cite{ww13}. Some desired properties of the new
definition were shown in a limited range of parameters and
conjectured for a larger range. The conjectures have been proved
soon in \cite{fl13,sb13}. The authors of \cite{mosog13} compared
the old and new forms of R\'{e}nyi's $\alpha$-divergences in the
context of quantum hypothesis testing. Hence, the proper choice
seems to be the traditional definition for $\alpha<1$ and the new
definition for $\alpha>1$ \cite{mosog13}. In the following, we
will focus an attention on the family of quantum $f$-divergences
defined by (\ref{qsfdef}).

Many results of quantum information theory are related to
monotonicity of the standard relative entropy
\cite{vedral02,nielsen}. The most general form of quantum
evolution is described by completely positive maps \cite{nielsen}.
Consider a linear map
$\Phi:{\>}\lnp(\hh)\rightarrow\lnp(\hh^{\prime})$ that takes
elements of $\lnp(\hh)$ to elements of $\lnp(\hh^{\prime})$. We
call a map positive, if it maps each positive operator to positive
one again. Let $\id^{\prime\prime}$ be the identity map on
$\lnp(\hh^{\prime\prime})$, where the space $\hh^{\prime\prime}$
is assigned to a reference system. The complete positivity implies
that the map $\Phi\otimes\id^{\prime\prime}$ is positive for any
dimension of $\hh^{\prime\prime}$. A completely positive map
$\Phi$ can be written as
\begin{equation}
\Phi(\htx)=\sum\nolimits_{n}{\htk_{n}{\,}\htx{\,}\htk_{n}^{\dagger}}
\ . \label{opsm}
\end{equation}
Here, the Kraus operators $\htk_{n}$ map the input space $\hh$ to
the output space $\hh^{\prime}$. When physical process is closed
and the probability is conserved, the map preserves the trace.
Trace-preserving completely positive maps (TPCP-maps) are called
quantum channels \cite{bengtsson,nielsen}. Entropic
characteristics of quantum channels were examined in
\cite{rfz10,rzf11}. In some respects, this treatment can be
extended with generalized entropic forms \cite{rast11a,rastcejp}.

Conditions for monotonicity of the quantum $f$-divergence are
obtained in \cite{hmp10}. Let $\Phi$ be a TPCP-map. If the
function $\xi\mapsto{f}(\xi)$ is operator convex on $[0;+\infty)$
then \cite{hmp10}
\begin{equation}
{D_{f}}{\bigl(\Phi(\hta){\big|\big|}{\,}\Phi(\htb)\bigr)}\leq{D}_{f}(\hta||\htb)
\ . \label{fdmon}
\end{equation}
The inequality (\ref{fdmon}) expresses monotonicity of
the quantum $f$-divergences. It will be very important in
studying quantum conditional entropies defined on base of the
$f$-divergences.

\section{Quantum conditional entropies}\label{sec3}

Conditional entropies are used in considering
multi-partite quantum systems. Let the state of a bipartite system
be described by density matrix $\htr_{AB}\in\lsp(\hh_{AB})$, where
$\hh_{AB}=\hh_{A}\otimes\hh_{B}$. The conditional von Neumann
entropy is defined as
\begin{equation}
H_{1}(\htr_{AB}|B):=H_{1}(\htr_{AB})-H_{1}(\htr_{B})
\ . \label{cvne}
\end{equation}
Here, the quantity $H_{1}(\htr)=-\tr(\htr\ln\htr)$ is the von
Neumann entropy of $\htr$, and the reduced density matrix $\htr_{B}$ is
obtained from $\htr_{AB}$ by tracing-out the space $\hh_{A}$. Some
useful relations between operators before and after partial
trace with applications to quantum entropies were obtained in
\cite{rastjst12}. Continuity of the conditional entropy
(\ref{cvne}) was considered in \cite{alf04}. The definition
(\ref{cvne}) has a lot of consequences \cite{nielsen}. Here, we recall 
two properties, since they will be referred to in the following. We have \cite{nielsen}
\begin{align}
&{H_{1}}{\bigl(|AB\rangle\langle{AB}|\big|{B}\bigr)}=-H_{1}(\htr_{B})
\ , \label{abpur}\\
&H_{1}(\htr_{A}\otimes\htr_{B}|B)=H_{1}(\htr_{A})
\ . \label{absep}
\end{align}
Thus, the conditional entropy (\ref{cvne}) reduces to minus the
entropy of $\htr_{B}$ in the case of pure states
$|AB\rangle\in\hh_{AB}$. It is herewith strictly negative for pure
entangled states. On the other hand, the conditional entropy
(\ref{cvne}) is nonnegative in the case of separable mixed
states. Such properties support a treatment of the quantity
(\ref{cvne}) as a quantum counterpart of the classical conditional
entropy.

Let us proceed to more general forms of quantum conditional
information. One of possible ways is a direct extension of the
definition (\ref{cvne}) with generalized entropies. It turned out
that an appropriate approach to obtaining more conditional
entropies can be based on quantum divergences. The key observation
is that the right-hand side of (\ref{cvne}) can be rewritten in an
alternative form \cite{rren05,krs09}:
\begin{equation}
H_{1}(\htr_{AB}|B):=
{-\inf}{\Bigl\{
{D_{1}}{\bigl(\htr_{AB}{\big|\big|}{\,}\pen_{A}\otimes\hts_{B}\bigr)}:
{\>}\hts_{B}\in\lsp(\hh_{B}),{\>}\tr(\hts_{B})=1\Bigr\}}
{\>}. \label{cvne1}
\end{equation}
This way to obtain quantum conditional entropies has been realized
for the so-called min-entropy and max-entropy in
\cite{rren05,krs09}. With minor modifications, this approach was
recently considered in the R\'{e}nyi case \cite{mdsft13}.

Following the idea of \cite{rren05,marco12}, the following
definition will be used. For the given $f$-divergence, we define
the associated quantum conditional entropy
\begin{equation}
H_{f}(\htr_{AB}|B):=
{-\inf}{\Bigl\{
{D_{f}}{\bigl(\htr_{AB}{\big|\big|}{\,}\pen_{A}\otimes\hts_{B}\bigr)}:
{\>}\hts_{B}\in\lsp(\hh_{B}),{\>}\tr(\hts_{B})\leq1\Bigr\}}
{\>}. \label{cvne2}
\end{equation}
As was mentioned above, the $f$-divergence is continuous in its
second entry (for details, see section 2 of \cite{hmp10}). We also
note that the optimization in (\ref{cvne2}) is taken over a
compact set. According to Weierstrass' theorem, the infimum is
finite and reached for at least one element of the set.

The optimization in (\ref{cvne2}) is taken over the set
of all sub-normalized states. There exist several reasons to allow
such states. For the min-entropy an optimization over the set of sub-normalized
states is equivalent to an optimization over normalized states
\cite{marco12}. In this case, the optimization problem is linear \cite{krs09}. Hence, 
convenient properties of the min-entropy follow. The given quantum system can
always be imagined as a part of some larger quantum systems. It is
natural to demand that the actual value of conditional entropies
be independent of any extension of the state space. The following argument 
demonstrates the physical meaning behind the sub-normalized states in (\ref{cvne2}).

Under conditions imposed on the function $\xi\mapsto{f}(\xi)$ (see
(\ref{fxlm}) and comments therein), the following property holds.
If the four positive semi-definite operators $\hta_{1}$, $\htb_{1}$, $\hta_{2}$,
$\htb_{2}$ obey
$\hta_{1}^{0}\vee\htb_{1}^{0}\perp\hta_{2}^{0}\vee\htb_{2}^{0}$
then \cite{hmp10}
\begin{equation}
{D_{f}}{\bigl(\hta_{1}+\hta_{2}{\big|\big|}{\,}\htb_{1}+\htb_{2}\bigr)}
=D_{f}(\hta_{1}||\htb_{1})+D_{f}(\hta_{2}||\htb_{2})
\ . \label{fdit}
\end{equation}
For arbitrary density operator $\htr_{AB}$, we have (see lemma
B.4.1 in \cite{rren05})
\begin{equation}
\ron(\htr_{AB})\subseteq\ron(\htr_{A})\otimes\ron(\htr_{B})
\ , \label{lb41}
\end{equation}
where $\htr_{A}=\tr_{B}(\htr_{AB})$ and
$\htr_{B}=\tr_{A}(\htr_{AB})$ are the reduced density matrices. Let two real numbers 
$\mu,\bar{\mu}\in[0;1]$ satisfy $\mu+\bar{\mu}\leq1$. Let the density matrices $\hto_{B}$ 
and $\htop_{B}$ be such that $\tr(\hto_{B})=\tr(\htop_{B})=1$, $\hto_{B}\in\lsp{\bigl(\ron(\htr_{B})\bigr)}$, and
$\htop_{B}^{0}\perp\htr_{B}^{0}$. We consider sub-normalized states
of the form
\begin{equation}
\hts_{B}^{\prime}=\mu{\,}\hto_{B}+{\bar{\mu}}{\,}\htop_{B}
\ . \label{sbsbp}
\end{equation}
Using the property (\ref{fdit}), for the case $f(0)=0$ one gets
\begin{equation}
{D_{f}}{\bigl(\htr_{AB}{\big|\big|}{\,}\pen_{A}\otimes\hts_{B}^{\prime}\bigr)}=
{D_{f}}{\bigl(\htr_{AB}{\big|\big|}{\,}\pen_{A}\otimes\mu{\,}\hto_{B}\bigr)}
\ , \label{ommu}
\end{equation}
since $D_{f}(\hta_{2}||\htb_{2})$ is zero for $\hta_{2}=\niz$. In the right-hand side of (\ref{ommu}), the state $\mu{\,}\hto_{B}$ 
has the trace $\mu$. Except for $\bar{\mu}=0$, this state is strictly sub-normalized. This reason shows that
sub-normalized states naturally occur in approaching the notion of
conditional entropy in terms of quantum divergences. 

We demand that the actual value of quantum conditional entropies be
independent of a choice of considered Hilbert space. Of course,
our consideration should involve at least the subspace
$\ron(\htr_{B})$ of $\hh_{B}$. This point is formally posed as follows. Let us treat the system $B$ 
as a particular case of extended quantum system $\wb$ with the Hilbert space $\hh_{\wb}=\hh_{B}\oplus\hk$. 
For all density matrices $\htr_{AB}$ on $\hh_{AB}$ and arbitrary finite space $\hk$, we formulate the condition
\begin{equation}
H_{f}(\htr_{AB}|B)=H_{f}(\htr_{AB}|\wb)	
\ . \label{hfwb}
\end{equation}
It turns out that the claim (\ref{hfwb}) can be provided by imposing some conditions on the used function $f$. 
We have the following statement. 

\newtheorem{tt0}{Theorem}
\begin{tt0}\label{tem0}
Let twice continuously differentiable function $\xi\mapsto{f}(\xi)$ be operator convex on
$[0;+\infty)$, let the limit (\ref{fxlm}) exist in $[-\infty;+\infty]$, and let $f(0)=0$. Then the quantum conditional entropy
(\ref{cvne2}) is represented as
\begin{equation}
H_{f}(\htr_{AB}|B)=
{-\inf}{\Bigl\{
{D_{f}}{\bigl(\htr_{AB}{\big|\big|}{\,}{\textup{\pen}}_{A}\otimes\hts_{B}\bigr)}:
{\>}\hts_{B}\in\lsp{\bigl(\ron(\htr_{B})\bigr)},{\>}\tr(\hts_{B})=1\Bigr\}}
{\>}. \label{cvne22}
\end{equation}
\end{tt0}

{\bf Proof.}
We aim to show that the definition (\ref{cvne2}) is
actually reduced to (\ref{cvne22}). We first note that the
optimization in (\ref{cvne2}) can be taken over states of the form
(\ref{sbsbp}). This claim follows from the monotonicity property.
Let us consider the TPCP-map $\Phi_{AB}$ with two Kraus operators
$\pen_{A}\otimes\htr_{B}^{0}$ and
$\pen_{A}\otimes{\bigl(\pen_{B}-\htr_{B}^{0}\bigr)}$. Due to
(\ref{lb41}), the state $\htr_{AB}$ is not altered by the map
$\Phi_{AB}$. For any sub-normalized state
$\hts_{B}\in\lsp(\hh_{B})$, we further obtain
\begin{equation}
\Phi_{AB}{\bigl(\pen_{A}\otimes\hts_{B}\bigr)}=\pen_{A}\otimes\hts_{B}^{\prime}
\ , \qquad
\mu={\tr}{\bigl(\htr_{B}^{0}{\,}\hts_{B}\bigr)}
\ , \qquad
{\bar{\mu}}=\tr(\hts_{B})-\mu
\ . \label{mubmu}
\end{equation}
Using the relations (\ref{fdmon}) and (\ref{ommu}), we
immediately obtain the inequality
\begin{equation}
{D_{f}}{\bigl(\htr_{AB}{\big|\big|}{\,}\pen_{A}\otimes\mu{\,}\hto_{B}\bigr)}
\leq{D_{f}}{\bigl(\htr_{AB}{\big|\big|}{\,}{\pen}_{A}\otimes\hts_{B}\bigr)}
\ , \label{monpr}
\end{equation}
in which $\hto_{B}\in\lsp{\bigl(\ron(\htr_{B})\bigr)}$ and
$\tr(\hto_{B})=1$. Hence, the question is actually reduced to
minimization over sub-normalized states on the subspace
$\ron(\htr_{B})$. Using item 2.3 of the paper \cite{hmp10}, the
left-hand side of (\ref{monpr}) can be rewritten as
\begin{equation}
\sum_{a\in\spc(\htr_{AB})}{{\left(
\sum_{b\in\spc(\hto_{B})\setminus\{0\}}{\mu{b}{\,}f{\left(\frac{a}{\mu{b}}\right)}
{\,}{\tr}{\bigl(\htp_{a}(\pen_{A}\otimes\htpi_{b})
\bigr)}}
+a{\,}\ell(f){\>}{\tr}{\bigl(\htp_{a}(\pen_{A}\otimes\htpi_{0})\bigr)}
\right)}}
{\,}. \label{qsf23}
\end{equation}
Here, the projectors $\htpi_{b}$ correspond to the eigenvalues of
$\hto_{B}$. If the twice continuously differentiable function $\xi\mapsto{f}(\xi)$ is
operator convex and $f(0)=0$, then the function
$\xi\mapsto\xi^{-1}f(\xi)$ is operator monotone (see, e.g.,
Corollary V.3.11 of the book \cite{bhatia97}).
Hence, the quantity (\ref{qsf23}) does not increase
with growth of $\mu$.  To reach the infimum, we should set $\mu$
to its maximum, i.e., to $\mu=1$. Combining this with (\ref{cvne2}) gives the claim (\ref{cvne22}). 
$\blacksquare$

Under the conditions imposed on the function $\xi\mapsto{f}(\xi)$, the
quantum conditional entropy (\ref{cvne2}) satisfies the properties (\ref{cvne22}) and (\ref{hfwb}). In other words, its
value is independent of any extension of the used Hilbert space.
The first condition is that the function should be operator
convex. Then the $f$-divergence is monotone. We also demand that
the function $\xi\mapsto\xi^{-1}f(\xi)$ be twice differentiable and
$f(0)=0$. One of important examples  is written as
\begin{equation}
f_{\alpha}(\xi)=\frac{\xi^{\alpha}-\xi}{\alpha-1}
\ , \label{falp0}	
\end{equation}
including $f_{1}(\xi)=\xi\ln\xi$.  The monotonicity property
(\ref{fdmon}) is one of the key steps in the proof of Theorem
\ref{tem0}. The monotonicity of the $f$-divergence implies that important
properties of the conditional entropy (\ref{cvne2}) hold. In the
next section, we examine this question in more detail. For
instance, we will show that the quantity (\ref{cvne2}) shares some
properties similar to the results (\ref{abpur}) and
(\ref{absep}) for the standard conditional entropy (\ref{cvne}).
In this sense, the considered extension differs from a generalized
conditional entropy determined by a local measurement
\cite{giro13}. The latter is always nonnegative and related to the
quantum discord.

Let us discuss briefly computability of the quantum conditional
entropy (\ref{cvne2}). The minimization of the $f$-divergence over
$\hts_{B}$ in (\ref{cvne22}) is a nonlinear optimization problem
with non-commuting variables. In general, such problems are
sufficiently difficult. Even in the commutative case, various bounds
on the $f$-divergences may be useful \cite{drag03}. On the other
hand, under some circumstances the quantum conditional entropy
(\ref{cvne2}) is relatively easy to calculate or estimate
analytically. For instance, more explicit relations can be
obtained in the case of partly classical states. In more details,
we will discuss results of such a kind in the next sections.
Sometimes, an explicit form of the function $\xi\mapsto{f}(\xi)$
is required. As a significant particular example, we will consider
the Tsallis case.

\section{Some basic properties}\label{sec4}

In this section, we consider basic properties of quantum
conditional entropies defined in terms of the $f$-divergences. In
general, we are always interested in some bounds, which describe a
range of possible values of studied quantities. Using arguments
from the proof of Theorem \ref{tem0}, we can obtain an upper bound
on the conditional entropy in terms of $\htr_{AB}$. A simple lower
bound easily follows from the definition of the conditional
entropy. The following statement takes place.

\newtheorem{tt01}[tt0]{Theorem}
\begin{tt01}\label{tem01}
Let the function $\xi\mapsto{f}(\xi)$ satisfy all the
preconditions of Theorem \ref{tem0}. Then the quantum conditional
entropy satisfies
\begin{equation}
-d_{B}^{-1}{\tr}{\bigl(f(d_{B}\htr_{AB})\bigr)}\leq
H_{f}(\htr_{AB}|B)\leq{-\tr}{\bigl(f(\htr_{AB})\bigr)}
\ . \label{dsdp2}
\end{equation}
\end{tt01}

{\bf Proof.} Let $\hto_{B}\in\lsp(\hh_{B})$ be normalized state
such that
\begin{equation}
H_{f}(\htr_{AB}|B)=
{-D_{f}}{\bigl(\htr_{AB}{\big|\big|}{\,}\pen_{A}\otimes\hto_{B}\bigr)}
\ . \label{chruf1}
\end{equation}
The upper bound is based on non-decreasing of the function
$\xi\mapsto\xi^{-1}f(\xi)$. Taking the expression (\ref{qsf23})
with $\mu=1$ and already without the second part, we get
\begin{equation}
{D_{f}}{\bigl(\htr_{AB}{\big|\big|}{\,}\pen_{A}\otimes\hto_{B}\bigr)}\geq
{D_{f}}{\bigl(\htr_{AB}{\big|\big|}{\,}\pen_{A}\otimes\pen_{B}\bigr)}
\ . \label{dsdp}
\end{equation}
Indeed, eigenvalues of the normalized $\hto_{B}$ do not exceed
$1$. It easily follows from (\ref{fabpq}) and (\ref{qsfdef}) that
\begin{equation}
{D_{f}}{\bigl(\htr_{AB}{\big|\big|}{\,}\pen_{A}\otimes\pen_{B}\bigr)}
={\tr}{\bigl(f(\htr_{AB})\bigr)}
\ . \label{dsdp1}
\end{equation}
Combining the inequalities (\ref{dsdp}) and (\ref{dsdp1}) with
(\ref{chruf1}) finally gives the upper bound. Since $\hto_{B}$ is
optimizing density matrix, we further write
\begin{equation}
{D_{f}}{\bigl(\htr_{AB}{\big|\big|}{\,}\pen_{A}\otimes\hto_{B}\bigr)}\leq
{D_{f}}{\bigl(\htr_{AB}{\big|\big|}{\,}\pen_{A}\otimes\htv_{*B}\bigr)}
\ , \label{ovvb}
\end{equation}
where the completely mixed state $\htv_{*B}=d_{B}^{-1}\pen_{B}$.
The operator $\pen_{A}\otimes\htv_{*B}$ has the eigenvalue
$d_{B}^{-1}$ with multiplicity $d_{A}d_{B}$ and the projector
$\pen_{A}\otimes\pen_{B}$. Reversing the sign in the right-hand
side of (\ref{ovvb}), we merely reduce it to the left-hand side of
(\ref{dsdp2}). $\blacksquare$

Note that the proof of Theorem \ref{tem01} tacitly uses
(\ref{cvne22}), whence all the conditions on the function
$\xi\mapsto{f}(\xi)$ are realized. In some interesting cases, the
lower and upper bounds of Theorem \ref{tem01} can be expressed in
terms of a generalized entropy of state $\htr_{AB}$. In the next
section, we will consider an important case of the Tsallis
entropies.

Under certain conditions of the used function, we will show the
conditional entropy (\ref{cvne2}) to be strictly negative for pure
entangled states. In this sense, the quantity (\ref{cvne2})
succeeds the relevant property (\ref{abpur}) of the standard
conditional entropy. Let twice continuously differentiable function $f(\xi)$ be operator convex 
and $f(0)=f(1)=0$. As was mentioned above, the function
$\xi\mapsto\xi^{-1}f(\xi)$ is then non-decreasing. To estimate the
infimum (\ref{cvne22}), we focus on finite-valued cases of the
$f$-divergence. For a pure state $|AB\rangle\in\hh_{AB}$, we write
\begin{equation}
{D_{f}}{\Bigl(|AB\rangle\langle{AB}|{\,}\Big|\Big|{\,}\pen_{A}\otimes\hts_{B}\Bigr)}=
\sum_{b\in\spc(\hts_{B})\setminus\{0\}} b{\,}f{\left(\frac{1}{b}\right)}
{\,}\tr(\htr_{B}\htpi_{b})
\ . \label{dfab}
\end{equation}
Here, we used $f(0)=0$ and the term
$\langle{AB}|\pen_{A}\otimes\htpi_{b}|AB\rangle=\tr(\htr_{B}\htpi_{b})$,
which follows from the Schmidt decomposition of $|AB\rangle$ and
properties of the partial trace. In the considered case, we can
rewrite (\ref{dfab}) merely as
${\tr}{\left(\htr_{B}{\,}\hts_{B}{\,}f(\hts_{B}^{-1})\right)}$.
Let us take $\hts_{B}$ as the completely mixed state on
$\ron(\htr_{B})$, which reads $\sch_{AB}^{-1}{\,}\htr_{B}^{0}$ in
terms of the Schmidt number $\sch_{AB}=\tr(\htr_{B}^{0})$ of the
state $|AB\rangle$. We then obtain from (\ref{dfab}) a two-sided
estimate
\begin{equation}
0\leq\underset{\tr(\hts_{B})=1}{\inf}{D_{f}}{\Bigl(|AB\rangle\langle{AB}|{\,}\Big|\Big|{\,}\pen_{A}\otimes\hts_{B}\Bigr)}
\leq\frac{f(\sch_{AB})}{\sch_{AB}}
\ . \label{ubin}
\end{equation}
Here, the zero bound follows as the function
$b\mapsto{b}{\,}f(1/b)$ is non-increasing, whence
$b{\,}f(1/b)\geq{f}(1)=0$ for $b\leq1$. Then the quantum
conditional entropy obeys
\begin{equation}
-{\,}\frac{f(\sch_{AB})}{\sch_{AB}}
\leq{H_{f}}{\bigl(|AB\rangle\langle{AB}|\big|{B}\bigr)}\leq0
\ . \label{hbin}
\end{equation}
Thus, the quantum conditional entropy (\ref{cvne2}) of any
bipartite pure state is not positive. We have arrived at this
claim due to conditions that the used function is operator convex
and $f(0)=f(1)=0$. We can improve the upper bound of (\ref{hbin}),
when the function $\xi\mapsto\xi^{-1}f(\xi)$ is additionally
convex (for a concrete example, see the next section). Indeed, the
function $b\mapsto{b}{\,}f(1/b)$ then becomes also convex.
Combining this with (\ref{dfab}) further gives
\begin{equation}
{D_{f}}{\Bigl(|AB\rangle\langle{AB}|{\,}\Big|\Big|{\,}\pen_{A}\otimes\hts_{B}\Bigr)}
\geq\tr(\htr_{B}\hts_{B}){\,}{f}{\left(\tr(\htr_{B}\hts_{B})^{-1}\right)}
\ . \label{dfab1}
\end{equation}
Here, we used the Jensen inequality and the fact that
probabilities $\tr(\htr_{B}\htpi_{b})$ are summarized to $1$. With
$\tr(\hts_{B})=1$, we clearly have
$\tr(\htr_{B}\hts_{B})\leq\|\htr_{B}\|_{\infty}$ (for a positive
operator, its spectral norm is merely the maximum of eigenvalues).
As the function $b\mapsto{b}{\,}f(1/b)$ is non-increasing, the
right-hand side of (\ref{dfab1}) is not less than
$\|\htr_{B}\|_{\infty}{\,}{f}{\left(\|\htr_{B}\|_{\infty}^{-1}\right)}$.
Hence, the quantum conditional entropy of a bipartite pure state
is bounded from above as
\begin{equation}
{H_{f}}{\bigl(|AB\rangle\langle{AB}|\big|{B}\bigr)}\leq
-{\,}\|\htr_{B}\|_{\infty}{\,}{f}{\left(\|\htr_{B}\|_{\infty}^{-1}\right)}
\ . \label{hbin1}
\end{equation}
We have
$\|\htr_{B}\|_{\infty}{\,}{f}{\left(\|\htr_{B}\|_{\infty}^{-1}\right)}>f(1)=0$,
when $\|\htr_{B}\|_{\infty}<1$ and the function is not constant.
The right-hand side of (\ref{hbin1}) and, therefore, the
conditional entropy are strictly negative for pure entangled
states. If Schmidt coefficients of the state $|AB\rangle$ are all
equal, then $\htr_{B}=\sch_{AB}^{-1}{\,}\htr_{B}^{0}$ and
$\|\htr_{B}\|_{\infty}=\sch_{AB}^{-1}$. Then the right-hand side
of (\ref{hbin1}) coincides with the left-hand side of (\ref{hbin}).
Thus, the lower bound of the relation (\ref{hbin}) is saturated
for such states, including maximally mixed states.

Thus, under certain conditions the presented generalization
(\ref{cvne2}) succeeds one of the basic properties of the standard
conditional entropy. In the next section, we will also exemplify
that the quantity (\ref{cvne2}) obeys the property quite similar
to (\ref{absep}). The following important property of quantum
conditional entropies is related to data processing.

\newtheorem{tt1}[tt0]{Theorem}
\begin{tt1}\label{tem1}
Let $\Psi_{B}:{\>}\lnp(\hh_{B})\rightarrow\lnp(\hh_{B^{\prime}})$
be a TPCP-map, and let $\id_{A}$ be the identity map on
$\lnp(\hh_{A})$. If the function $\xi\mapsto{f}(\xi)$ obeys all the
preconditions of Theorem \ref{tem0}, then
\begin{equation}
H_{f}(\htr_{AB}|B)\leq
H_{f}(\htr_{AB^{\prime}}|B^{\prime})
\ , \label{hbbp0}
\end{equation}
where $\htr_{AB^{\prime}}=\id_{A}\otimes\Psi_{B}(\htr_{AB})$.
\end{tt1}

{\bf Proof.} Let $\hto_{B}\in\lsp(\hh_{B})$ be normalized state
such that the formula (\ref{chruf1}) holds. We first note that
\begin{equation}
\id_{A}\otimes\Psi_{B}\bigl(\pen_{A}\otimes\hto_{B}\bigr)=\pen_{A}\otimes\Psi_{B}(\hto_{B})
\ . \label{apso}
\end{equation}
The monotonicity property (\ref{fdmon}) then gives
\begin{equation}
{D_{f}}{\bigl(\htr_{AB^{\prime}}{\big|\big|}{\,}\pen_{A}\otimes\Psi_{B}(\hto_{B})\bigr)}
\leq
{D_{f}}{\bigl(\htr_{AB}{\big|\big|}{\,}\pen_{A}\otimes\hto_{B}\bigr)}
\ . \label{fdmon1}
\end{equation}
Since the output $\Psi_{B}(\hto_{B})$ is a density matrix on
$\hh_{B^{\prime}}$, the definition (\ref{cvne2}) implies
\begin{equation}
{-D_{f}}{\bigl(\htr_{AB^{\prime}}{\big|\big|}{\,}\pen_{A}\otimes\Psi_{B}(\hto_{B})\bigr)}
\leq{H}_{f}(\htr_{AB^{\prime}}|B^{\prime})
\ . \label{hbbp1}
\end{equation}
Rearranging the two sides of (\ref{fdmon1}) with the relevant
sign, the formulas (\ref{chruf1}) and (\ref{hbbp1}) provide the
claim (\ref{hbbp0}). $\blacksquare$

For the new version of conditional R\'{e}nyi's entropy, an
inequality of the form (\ref{hbbp0}) has been obtained in
\cite{mdsft13}. We see that this inequality holds for an entire
family of quantum conditional entropies based on the quantum
$f$-divergences. An immediate corollary of the property
(\ref{hbbp0}) is posed as follows. Let $\htr_{ABC}$ be density
matrix on $\hh_{ABC}=\hh_{A}\otimes\hh_{B}\otimes\hh_{C}$. Then we
have
\begin{equation}
H_{f}(\htr_{ABC}|BC)\leq
H_{f}(\htr_{AB}|B)
\ , \label{hbbp00}
\end{equation}
where the density matrix $\htr_{AB}$ is obtained from $\htr_{ABC}$
by tracing-out $\hh_{C}$. To check (\ref{hbbp00}), we use the map
$\id_{A}\otimes\id_{B}\otimes\Psi_{C}$ with $\Psi_{C}$ defined for
all $|c_{1}\rangle,|c_{2}\rangle\in\hh_{C}$ as
\begin{equation}
{\Psi_{C}}{\bigl(|c_{2}\rangle\langle{c}_{1}|\bigr)}
=\langle{c}_{1}|c_{2}\rangle
\ . \label{psc12}
\end{equation}
By linearity, the definition (\ref{psc12}) is extended to
all elements of $\lnp(\hh_{C})$. It is clear that the map
$\Psi_{C}$ is a TPCP-map such that
\begin{equation}
\id_{A}\otimes\id_{B}\otimes\Psi_{C}(\htr_{ABC})=\htr_{AB}
\ . \label{ptr}
\end{equation}
According to (\ref{hbbp00}), conditioning on more can only reduce
the entropy. It is of interest to obtain an inequality in opposite
direction. Apparently, an explicit form of the used function
$\xi\mapsto{f}(\xi)$ is required here. For the conditional Tsallis
entropy, this issue will be considered below in Sect. \ref{sec5}.

It is of interest to analyze properties of the quantum conditional
entropy with respect to partly classical states. Let us take a
collection of density matrices $\htr_{Ay}\in\lsp(\hh_{A})$ such
that projectors $\htr_{Ay}^{0}$ are all mutually orthogonal, i.e.,
\begin{equation}
\htr_{Ax}^{0}\perp\htr_{Ay}^{0}
\qquad (x\neq{y})
\ , \label{xyprp}
\end{equation}
for all pairs $x\neq{y}$. To probability distribution $\{p_{y}\}$,
we assign a density matrix
\begin{equation}
\htr_{AY}=\sum\nolimits_{y} p_{y}{\,}\htr_{Ay}
\ . \label{rhay}
\end{equation}
It is said that the state (\ref{rhay}) has a classical $Y$-register
\cite{mdsft13}. The following statement takes place.

\newtheorem{tt2}[tt0]{Theorem}
\begin{tt2}\label{tem2}
Let density matrix
$\htr_{ABY}\in\lsp{\bigl(\hh_{A}\otimes\hh_{B}\bigr)}$ be given in
the form
\begin{equation}
\htr_{ABY}=\sum\nolimits_{y} p_{y}{\,}\htr_{ABy}
\ , \label{rhaby}
\end{equation}
in which $\htr_{ABx}^{0}\perp\htr_{ABy}^{0}$ for all $x\neq{y}$.
Suppose also that $\htr_{Bx}^{0}\perp\htr_{By}^{0}$ for all
$x\neq{y}$, where $\htr_{By}=\tr_{A}(\htr_{ABy})$ is the partial
trace over $\hh_{A}$. Let the function $\xi\mapsto{f}(\xi)$
satisfy all the preconditions of Theorem \ref{tem0}. Then the
quantum conditional entropy satisfies
\begin{equation}
\sum\nolimits_{y} p_{y}{\,}H_{f}(\htr_{ABy}|B)\leq{H}_{f}(\htr_{ABY}|BY)
\ . \label{clry}
\end{equation}
\end{tt2}

{\bf Proof.} We recall already mentioned properties of the
$f$-divergences. The first is
expressed by the formula (\ref{fdit}). For
$\lambda\in[0;+\infty)$, we also use the homogeneity (\ref{fhom}).
By (\ref{lb41}), for each $\htr_{ABy}$ we have
$\ron(\htr_{ABy})\subseteq\ron(\htr_{Ay})\otimes\ron(\htr_{By})$
in terms of the partial traces $\htr_{Ay}=\tr_{B}(\htr_{ABy})$ and
$\htr_{By}=\tr_{A}(\htr_{ABy})$. So, we can treat $\htr_{ABy}$ as
a positive operator on $\hh_{A}\otimes\ron(\htr_{By})$. Let
$\hto_{By}\in\lsp{\bigl(\ron(\htr_{By})\bigr)}$ be density
operator such that
\begin{equation}
H_{f}(\htr_{ABy}|B)=
{-D_{f}}{\bigl(\htr_{ABy}{\big|\big|}{\,}\pen_{A}\otimes\hto_{By}\bigr)}
\ . \label{chruf2}
\end{equation}
Introducing the state $\hto_{BY}=\sum_{y}p_{y}{\,}\hto_{By}$, we
directly obtain
\begin{align}
{D_{f}}{\bigl(\htr_{ABY}{\big|\big|}{\,}\pen_{A}\otimes\hto_{BY}\bigr)}&=
\sum\nolimits_{y}
{D_{f}}{\bigl(p_{y}{\,}\htr_{ABy}{\big|\big|}{\,}\pen_{A}\otimes{p}_{y}{\,}\hto_{By}\bigr)}
\label{cfdit}\\
&=\sum\nolimits_{y} p_{y}{\,}
{D_{f}}{\bigl(\htr_{ABy}{\big|\big|}{\,}\pen_{A}\otimes\hto_{By}\bigr)}
\ . \label{cfhom}
\end{align}
Since $\htr_{Bx}^{0}\perp\htr_{By}^{0}$ for $x\neq{y}$, the
subspaces $\hh_{A}\otimes\ron(\htr_{Bx})$ and
$\hh_{A}\otimes\ron(\htr_{By})$ contains mutually orthogonal
vectors. Here, the step (\ref{cfdit}) is due to (\ref{fdit}), and
the step (\ref{cfhom}) is due to (\ref{fhom}). According to
(\ref{cvne22}), we also have
\begin{equation}
{-D_{f}}{\bigl(\htr_{ABY}{\big|\big|}{\,}\pen_{A}\otimes\hto_{BY}\bigr)}
\leq{H}_{f}(\htr_{ABY}|BY)
\ . \label{dflc}
\end{equation}
Rearranging the terms in (\ref{cfhom}) with the corresponding
sign, we use (\ref{chruf2}) and (\ref{dflc}) to complete the
proof. $\blacksquare$

The result (\ref{clry}) relates the conditional entropy
$H_{f}(\htr_{ABY}|BY)$ with the conditional entropies of
particular states $\htr_{ABy}$. For the case of conditional
R\'{e}nyi entropies, a relation of such a kind has been derived in
\cite{mdsft13}. Moreover, the writers of \cite{mdsft13} gave an
exact relation. Below, we will obtain a similar exact relation for
the Tsallis case.

\section{Quantum conditional entropies of Tsallis type}\label{sec5}

Let us consider the function (\ref{falp0})
with positive $\alpha\neq1$. In classical regime, the formula
(\ref{cfdf}) leads to the Tsallis relative entropy
\begin{equation}
S_{\alpha}(p{||}q):=\frac{1}{1-\alpha}{\>}{\left(1-\sum\nolimits_{x}p_{x}^{\alpha}{\,}q_{x}^{1-\alpha}\right)}
=-\sum\nolimits_{x}p_{x}{\>}{\ln_{\alpha}}{\left(\frac{q_{x}}{p_{x}}\right)}
{\>}. \label{srtdef2}
\end{equation}
This quantity was considered in \cite{borland}. In
(\ref{srtdef2}), the $\alpha$-logarithm is defined as
\begin{equation}
\ln_{\alpha}(\xi):=\frac{\xi^{1-{\alpha}}-1}{1-{\alpha}}
\ , \label{aldf}
\end{equation}
where $\alpha>0\neq1$ and $\xi>0$. As the function
$f_{\alpha}(\xi)$ is convex and $f_{\alpha}(1)=0$, we have
$S_{\alpha}(p{||}q)\geq0$ due to (\ref{cfin}). In the limit
$\alpha\to1$, the $\alpha$-logarithm is reduced to the usual one.
The quantity (\ref{srtdef2}) then gives the standard relative
entropy of probability distributions. Some bounds on the classical
relative entropy (\ref{srtdef2}) were obtained in
\cite{ics66,ggil}. In quantum regime, this question has been
addressed in \cite{rastmpag}.

Using (\ref{qsfdep}) with the function $f_{\alpha}(\xi)$, we
obtain the quantum version of Tsallis' relative entropy. For
$\alpha\in(1;+\infty)$, the Tsallis $\alpha$-divergence is defined
as
\begin{equation}
D_{\alpha}(\hta||\htb):=
\left\{
\begin{array}{ll}
\frac{1}{\alpha-1}\Bigl(\tr(\hta^{\alpha}\htb^{1-\alpha})-\tr(\hta)\Bigr){\,},
& {\mathrm{if{\ }}}\ron(\hta)\subseteq\ron(\htb){\>}, \\
+\infty{\>}, & {\mathrm{otherwise}}{\>}.
\end{array}
\right.
\label{qendf}
\end{equation}
For $\alpha\in(0;1)$, the first entry of (\ref{qendf}) is always
suitable. According to (\ref{aldf}), we have
$f_{\alpha}(\xi)=\xi^{\alpha}\ln_{\alpha}(\xi)$. The standard
relative entropy (\ref{qendf1}) is reached from (\ref{qendf}) in
the limit $\alpha\to1$. The writers of \cite{akr13} recently
proposed a ``sandwiched'' relative Tsallis entropy connected with
the new definition of R\'{e}nyi's divergence \cite{mdsft13,wwy13}.
The corresponding conditional  Tsallis form has been defined and
used in identifying entanglement \cite{akr13}. When the subsystem
density matrix is a maximally mixed state, this ``sandwiched''
conditional Tsallis entropy gives the nonadditive conditional
entropy proposed by Abe and Rajagopal \cite{ar01}. Such an
approach is beyond the scope of the present work.

The function $z\mapsto{z}^{\alpha}$ is operator concave on
$[0;+\infty)$ for $0\leq\alpha\leq1$ and operator convex on
$[0;+\infty)$ for $1\leq\alpha\leq2$ (see, e.g., items
4.2.3 and 1.5.8 in \cite{bhatia07}). Hence, the function
$f_{\alpha}(\xi)=\bigl(\xi^{\alpha}-\xi\bigr)/(\alpha-1)$ is
operator convex for $\alpha\in(0;2]$ and $\alpha\neq1$. Combining
this with the inequality (\ref{fdmon}) then gives
\begin{equation}
{D_{\alpha}}{\bigl(\Phi(\hta){\big|\big|}{\,}\Phi(\htb)\bigr)}\leq{D}_{\alpha}(\hta||\htb)
\ . \label{fdmona}
\end{equation}
For $\alpha\in(0;2]$, the quantum Tsallis divergence is monotone
under the action of trace-preserving completely positive maps.
Monotonicity of the divergence leads to a lot of properties of the
conditional Tsallis entropy. Since $f_{\alpha}(0)=0$, we apply
(\ref{cvne22}) and obtain the quantum conditional entropy
\begin{equation}
H_{\alpha}(\htr_{AB}|B):=
{-\inf}{\Bigl\{
{D_{\alpha}}{\bigl(\htr_{AB}{\big|\big|}{\,}\pen_{A}\otimes\hts_{B}\bigr)}:
{\>}\hts_{B}\in\lsp(\hh_{B}),{\>}\tr(\hts_{B})=1\Bigr\}}
{\>}. \label{cvne3}
\end{equation}
In this definition, we will assume $\alpha\in(0;2]$, whence the
Tsallis $\alpha$-divergence (\ref{qendf}) is monotone.

Let us discuss briefly the conditional Tsallis entropy of pure
states. We first note that the function
\begin{equation}
\xi^{-1}f_{\alpha}(\xi)=\frac{\xi^{\alpha-1}-1}{\alpha-1}={-\ln_{\alpha}}{\left(\frac{1}{\xi}\right)}
\label{xial}
\end{equation}
is increasing and convex for $\alpha\in(0;2]$. Combining this with
(\ref{hbin}) and (\ref{hbin1}) gives a two-sided estimate
\begin{equation}
{\ln_{\alpha}}{\left(\frac{1}{\sch_{AB}}\right)}
\leq{H_{\alpha}}{\Bigl(|AB\rangle\langle{AB}|\Bigm|{B}\Bigr)}\leq
{\ln_{\alpha}}{\left(\|\htr_{B}\|_{\infty}\right)}
\ . \label{hbint}
\end{equation}
For entangled pure states, we have $\|\htr_{B}\|_{\infty}<1$ and
strictly negative value of the conditional Tsallis entropy. The
lower and upper bounds of the relation (\ref{hbint}) coincide in
the case, when Schmidt coefficients of the state $|AB\rangle$ are
all equal. Then the conditional Tsallis entropy is equal to
${\ln_{\alpha}}{\left(\sch_{AB}^{-1}\right)}$, including
${-\ln}{\left(\sch_{AB}\right)}$ for the usual case $\alpha=1$. We
further consider the conditional Tsallis entropy of separable
states. Using the representation (\ref{qendf}), for any
$\hts_{B}\in\lsp{\bigl(\ron(\htr_{B})\bigr)}$ with the same range
we write
\begin{align}
{D_{\alpha}}{\bigl(\htr_{A}\otimes\htr_{B}{\big|\big|}{\,}\pen_{A}\otimes\hts_{B}\bigr)}
&=\frac{1}{\alpha-1}{\>}\Bigl(\tr(\htr_{A}^{\alpha})\tr(\htr_{B}^{\alpha}\hts_{B}^{1-\alpha})-1\Bigr)
\nonumber\\
&=\tr(\htr_{A}^{\alpha})D_{\alpha}(\htr_{B}||\hts_{B})-H_{\alpha}(\htr_{A})
\ . \label{dabsp}
\end{align}
Here, the quantum Tsallis $\alpha$-entropy of normalized state
$\htr_{A}$ is defined as
\begin{equation}
H_{\alpha}(\htr_{A}):=\frac{1}{\alpha-1}{\,}\Bigl(\tr(\htr_{A}^{\alpha})-1\Bigr)
=-D_{\alpha}(\htr_{A}||\pen_{A})
\ . \label{had}
\end{equation}
Properties of the quantum Tsallis entropy are discussed in
\cite{raggio}. With normalized $\hts_{B}$, we have
$D_{\alpha}(\htr_{B}||\hts_{B})\geq0$ (see (\ref{rspt4}) and
related comments). That is, the infimum of (\ref{dabsp}) is equal
to $-H_{\alpha}(\htr_{A})$; it is clearly reached for
$\hts_{B}=\htr_{B}$. From (\ref{cvne3}), we finally obtain
\begin{equation}
H_{\alpha}(\htr_{A}\otimes\htr_{B}|B)=H_{\alpha}(\htr_{A})
\ . \label{tabsp}
\end{equation}
Thus, for $\alpha\in(0;2]$ the quantum conditional entropy
(\ref{cvne3}) succeeds the property (\ref{absep}) of the standard
conditional entropy (\ref{cvne}). It is certainly nonnegative in
the case of separable mixed states.

Applying (\ref{dsdp2}) with
$f_{\alpha}(\xi)=\bigl(\xi^{\alpha}-\xi\bigr)/(\alpha-1)$, we
obtain the corresponding lower and upper bounds:
\begin{equation}
d_{B}^{\alpha-1}H_{\alpha}(\htr_{AB})+{\ln_{\alpha}}{\left(d_{B}^{-1}\right)}\leq
H_{\alpha}(\htr_{AB}|B)\leq{H}_{\alpha}(\htr_{AB})
\ . \label{dsdp2t}
\end{equation}
These bounds are based on (\ref{cvne22}) and hold
for all $\alpha\in(0;2]$. In the standard case $\alpha=1$, these
bounds are merely reduced to
\begin{equation}
H_{1}(\htr_{AB})-\ln{d}_{B}\leq
H_{1}(\htr_{AB}|B)\leq{H}_{1}(\htr_{AB})
\ . \label{dsdp2s}
\end{equation}
Of course, this claim also follows from (\ref{cvne}) due to
$0\leq{H}_{1}(\htr_{B})\leq\ln{d}_{B}$. The result
(\ref{dsdp2t}) is an extension of (\ref{dsdp2s}) to conditional
Tsallis' entropies.

The results (\ref{hbbp0}) and (\ref{hbbp00}) are based on the
monotonicity property. They hold for the conditional Tsallis
$\alpha$-entropy of degree $\alpha\in(0;2]$. In particular, we
have 
\begin{equation}
H_{\alpha}(\htr_{ABC}|BC)\leq{H}_{\alpha}(\htr_{AB}|B)
\ . \label{neweq}	
\end{equation}
As was already mentioned, we can obtain an inequality in opposite
direction. Namely, we have the following statement.

\newtheorem{tt3}[tt0]{Theorem}
\begin{tt3}\label{tem3}
Let $\htr_{ABC}$ be density matrix on the product
$\hh_{ABC}=\hh_{A}\otimes\hh_{B}\otimes\hh_{C}$. For all
$\alpha\in(0;2]$, the quantum conditional $\alpha$-entropy
satisfies
\begin{equation}
H_{\alpha}(\htr_{ABC}|B)\leq{d}_{C}^{1-\alpha}H_{\alpha}(\htr_{ABC}|BC)+\ln_{\alpha}(d_{C})
\ , \label{chrul0}
\end{equation}
where $d_{C}$ is the dimensionality of $\hh_{C}$.
\end{tt3}

{\bf Proof.} We also suppose $\alpha\neq1$. Let
$\hto_{B}\in\lsp(\hh_{B})$ be normalized state such that
\begin{equation}
H_{\alpha}(\htr_{ABC}|B)=
{-D_{\alpha}}{\bigl(\htr_{AB}{\big|\big|}{\,}\pen_{A}\otimes\hto_{B}\otimes\pen_{C}\bigr)}
\ . \label{chrul1}
\end{equation}
Using the completely mixed state $\htv_{*C}=d_{C}^{-1}\pen_{C}$ on
$\hh_{C}$, we write
\begin{align}
{D_{\alpha}}{\bigl(\htr_{AB}{\big|\big|}{\,}\pen_{A}\otimes\hto_{B}\otimes\pen_{C}\bigr)}
&=\frac{1}{\alpha-1}{\>}
{\left[
d_{C}^{1-\alpha}{\,}{\tr}{\Bigl(\htr_{ABC}^{\alpha}(\pen_{A}\otimes\hto_{B}\otimes\htv_{*C})^{1-\alpha}\Bigr)}
-1 \right]}
\nonumber\\
&=d_{C}^{1-\alpha}{\,}{D_{\alpha}}{\bigl(\htr_{AB}{\big|\big|}{\,}\pen_{A}\otimes\hto_{B}\otimes\htv_{*C}\bigr)}
-\ln_{\alpha}(d_{C})
\ . \label{chrul2}
\end{align}
According to the definition (\ref{cvne3}), we also have
\begin{equation}
{-D_{\alpha}}{\bigl(\htr_{AB}{\big|\big|}{\,}\pen_{A}\otimes\hto_{B}\otimes\htv_{*C}\bigr)}
\leq{H}_{\alpha}(\htr_{ABC}|BC)
\ , \label{chrul3}
\end{equation}
since the product $\hto_{B}\otimes\htv_{*C}$ is a density matrix
on $\hh_{BC}=\hh_{B}\otimes\hh_{C}$. After substituting
(\ref{chrul2}) into (\ref{chrul1}), the relation (\ref{chrul3})
completes the proof for $\alpha\neq1$. The case $\alpha=1$ can be
resolved in a similar manner. $\blacksquare$

Analogous property has been proved for the conditional min-entropy
by Renner \cite{rren05} and for the new form of conditional
R\'{e}nyi's entropy in \cite{mdsft13}. This property is usually
referred to as the chain rule. Thus, we have derived the chain
rule (\ref{chrul0}) in the Tsallis formulation. Note that the
proof of Theorem \ref{tem3} tacitly uses the result
(\ref{cvne22}), whence the restriction $\alpha\in(0;2]$ occurs.

In conclusion, we consider the conditional Tsallis entropy of
partly classical states. The inequality (\ref{clry}) has been
derived for any function $\xi\mapsto{f}(\xi)$ that obeys the
preconditions of Theorem \ref{tem0}. When this function is given
explicitly, we could obtain an exact relation instead of
(\ref{clry}). For the Tsallis case, it is posed as follows.

\newtheorem{tt4}[tt0]{Theorem}
\begin{tt4}\label{tem4}
Let density matrix
$\htr_{ABY}\in\lsp{\bigl(\hh_{A}\otimes\hh_{B}\bigr)}$ be given by
(\ref{rhaby}), and let projectors $\htr_{ABy}^{0}$ and
$\htr_{By}^{0}$ obey the same preconditions as in Theorem
\ref{tem2}. For $\alpha\in(0;2]$, the conditional Tsallis entropy
satisfies
\begin{equation}
H_{\alpha}(\htr_{ABY}|BY)=
\frac{1}{1-\alpha}\left[
{\left(\sum\nolimits_{y}p_{y}
{\bigl(1+(1-\alpha)H_{\alpha}(\htr_{ABy}|B)\bigr)}^{1/\alpha}
\right)}^{\alpha}-1
\right]
{\,}. \label{clryt}
\end{equation}
\end{tt4}

{\bf Proof.} We first assume that $\alpha\neq1$. According to the
definition (\ref{cvne3}), we aim to find the infimum of
${D_{\alpha}}{\bigl(\htr_{ABY}{\big|\big|}{\,}\pen_{A}\otimes\hts_{BY}\bigr)}$
over all density matrices $\hts_{BY}$ on $\hh_{B}$. Following
\cite{mdsft13}, we first note that the optimization can be taken
over states of the form
\begin{equation}
\hts_{BY}^{\prime}=\sum\nolimits_{y} q_{y}{\,}\hts_{By}
\ . \label{rhabys}
\end{equation}
Here, the set $\{q_{y}\}$ is some probability distribution and
$\hts_{By}\in\lsp{\bigl(\ron(\htr_{By})\bigr)}$.
The reason is posed as follows. Adding zero probabilities to
$\{p_{y}\}$, we can always assume the set of projectors
$\htr_{By}^{0}$ to be complete in $\hh_{B}$, namely
\begin{equation}
\sum\nolimits_{y}\htr_{By}^{0}=\pen_{B}
\ . \label{coply}
\end{equation}
We now apply a TPCP-map $\Phi_{ABY}$ with Kraus operators
$\htk_{y}=\pen_{A}\otimes\htr_{By}^{0}$. Due to
$\htr_{Bx}^{0}\perp\htr_{By}^{0}$ for $x\neq{y}$, this map
transforms an entry $\pen_{A}\otimes\hts_{BY}$ with any
$\hts_{BY}$ into an entry $\pen_{A}\otimes\hts_{BY}^{\prime}$ with
a density matrix $\hts_{BY}^{\prime}$ of the form (\ref{rhabys}).
The state (\ref{rhaby}) remains unaltered, since
$\ron(\htr_{ABy})\subseteq\hh_{A}\otimes\ron(\htr_{By})$. By the
monotonicity (\ref{fdmona}), the divergence cannot increase under
the action of the map $\Phi_{ABY}$, as claimed. For the states
(\ref{rhaby}) and (\ref{rhabys}), immediate calculations give
\begin{equation}
{D_{\alpha}}{\bigl(\htr_{ABY}{\big|\big|}{\,}\pen_{A}\otimes\hts_{BY}^{\prime}\bigr)}=
S_{\alpha}(p{||}q)+\sum\nolimits_{y} p_{y}^{\alpha}{\,}q_{y}^{1-\alpha}
{D_{\alpha}}{\bigl(\htr_{ABy}{\big|\big|}{\,}\pen_{A}\otimes\hts_{By}\bigr)}
\ , \label{dsda}
\end{equation}
where we used (\ref{fdit}), (\ref{srtdef2}), and (\ref{qendf}).
The right-hand side of (\ref{dsda}) should be minimized over all
probability distributions $\{q_{y}\}$ and state collections
$\{\hts_{By}\}$. Let $\hto_{B}\in\lsp(\hh_{B})$ be density
operator such that the equality (\ref{chruf2}) actually holds. The
minimization over $\{\hts_{By}\}$ merely leads to substituting the
collection $\{\hto_{By}\}$. Indeed, the factor of each
$\alpha$-divergence in the right-hand side of (\ref{dsda}) is
nonnegative. Then we have
\begin{equation}
-H_{\alpha}(\htr_{ABY}|BY)={\inf}{\left\{\frac{1}{\alpha-1}
\left(\sum\nolimits_{y} p_{y}^{\alpha}{\,}q_{y}^{1-\alpha}t_{y}-1\right):
{\>}q_{y}\geq0,{\>}\sum\nolimits_{y}q_{y}=1
\right\}}
\ , \label{funte}
\end{equation}
where we fix the distribution $\{p_{y}\}$ and auxiliary parameters
\begin{equation}
t_{y}=1+(1-\alpha)H_{\alpha}(\htr_{ABy}|B)=
{\tr}{\Bigl(\htr_{ABy}^{\alpha}(\pen_{A}\otimes\hto_{By})^{1-\alpha}\Bigr)}
\ . \label{tedf}
\end{equation}
Clearly, these parameters are nonnegative. In the term
(\ref{funte}), therefore, we deal with a convex function of the
variables $q_{y}$. Indeed, its second derivative with respect to
$q_{x}$ reads
$\alpha{p}_{x}^{\alpha}{\,}t_{x}{\,}q_{x}^{-\alpha-1}$ and is
nonnegative for $\alpha>0$. The method of Lagrange multipliers is
a standard tool to fit problems of minimizing convex function
subject to convex constraints \cite{rockaf}. Applying this
technique to the task (\ref{funte}) leads to the answer
\begin{equation}
q_{x}=\left(\sum\nolimits_{y}t_{y}^{1/\alpha}p_{y}\right)^{-1}
t_{x}^{1/\alpha}p_{x}
\ , \label{qans}
\end{equation}
which takes into account the normalization condition. Substituting
the formulas (\ref{tedf}) and (\ref{qans}) and into the right-hand
side of (\ref{funte}) finally gives (\ref{clryt}). By inspection of
second derivatives, we see that the found point is a conditional
minimum. The proof in the standard case $\alpha=1$ is short. Using
(\ref{qendf1}), we immediately obtain the formula
\begin{equation}
{D_{1}}{\bigl(\htr_{ABY}{\big|\big|}{\,}\pen_{A}\otimes\hts_{BY}^{\prime}\bigr)}=
S_{1}(p{||}q)+\sum\nolimits_{y}p_{y}{\,}
{D_{1}}{\bigl(\htr_{ABy}{\big|\big|}{\,}\pen_{A}\otimes\hts_{By}\bigr)}
\ , \label{dsda1}
\end{equation}
which is also a limit case of the result (\ref{dsda}). Two
summands in the right-hand side of (\ref{dsda1}) are minimized
independently. Substituting the collection $\{\hto_{By}\}$
optimizes the second summand over $\{\hts_{By}\}$. The first
summand obeys $S_{1}(p{||}q)\geq0$ (see, e.g., theorem 11.1 in
\cite{nielsen}) and vanishes, when $q_{y}=p_{y}$ for all $y$.
Hence, we obtain the result
\begin{equation}
H_{1}(\htr_{ABY}|BY)=
\sum\nolimits_{y}p_{y}
{\,}H_{1}(\htr_{ABy}|B)
\ , \label{clryt1}
\end{equation}
which follows from (\ref{clryt}) in the limit $\alpha\to1$.
$\blacksquare$

The statement of Theorem \ref{tem4} gives an expression of the
conditional Tsallis $\alpha$-entropy in the case of partly
classical states. For the conditional R\'{e}nyi entropy, this
issue was already considered in \cite{mdsft13}. Note that the
expression (\ref{clryt1}) for $\alpha\to1$ follows from the
R\'{e}nyi formulation as well. As a particular case of
(\ref{clryt}), we obtain the result
\begin{equation}
H_{\alpha}(\htr_{AY}|Y)=
\frac{1}{1-\alpha}\left[
{\left(\sum\nolimits_{y}p_{y}
{\bigl(1+(1-\alpha)H_{\alpha}(\htr_{Ay})\bigr)}^{1/\alpha}
\right)}^{\alpha}-1
\right]
{\,}, \label{cbryt}
\end{equation}
in which conditioning is purely classical. Here, the restriction
$\alpha\in(0;2]$ is essential again. We see that the conditional
Tsallis entropy obeys some relations similarly to the conditional
R\'{e}nyi entropy.

\section{Conclusions}\label{sec6}

We have studied quantum conditional entropies defined in terms
of the $f$-divergences. This approach has previously been applied
to  the so-called min-entropy and max-entropy in
\cite{rren05,krs09}. Recently, it was considered for the R\'{e}nyi
case \cite{mdsft13}. We have applied the mentioned method to a
family of the quantum $f$-divergences. We gave explicit
conditions, when the notion of quantum conditional entropies is
well defined in the developed approach. This question is naturally
raised from the fact that any quantum system can always be
imagined as a part of larger quantum systems. The additional
conditions on $f$ are that it is twice continuously differentiable and $f(0)=0$. 
Together with the operator convexity, the conditions
imply non-decreasing of the function $\xi\mapsto\xi^{-1}f(\xi)$.
When other conditions are already satisfied, this property can
easily be reached.

Assuming these conditions, important properties of the introduced
conditional entropies have been discussed. In particular, the
presented quantity resembles properties of the standard
conditional entropy in both the cases of pure states and separable
states. We also derived simple lower and upper bounds on the
conditional entropy in terms of the corresponding density matrix.
A behavior with respect to data processing was considered. Quantum
conditional entropies of partly classical states were examined. We
have considered an especially important case of the quantum
conditional entropies of Tsallis type. In some respects, their
properties are similar to quantum conditional R\'{e}nyi entropies
recently studied in \cite{mdsft13}. The presented discussion is a
further development of the approach originally proposed in
\cite{rren05,krs09}.


\begin{thebibliography}{99}

\bibitem{vedral02}
V. Vedral:  {\em Rev.\ Mod.\ Phys.\/} {\bf 74}, 197--234 (2002).

\bibitem{petz08}
D. Petz: {\em Quantum Information Theory and Quantum Statistics\/}, Springer, Berlin 2008.  

\bibitem{bengtsson}
I. Bengtsson and K. \.{Z}yczkowski: {\em Geometry of Quantum States: An
Introduction to Quantum Entanglement\/}, Cambridge University Press, Cambridge 2006.

\bibitem{hmp10}
F. Hiai, M. Mosonyi, D. Petz and C. B\'{e}ny: {\em Rev.\ Math.\ Phys.\/} {\bf 23}, 691--747 (2011).

\bibitem{ics67}
I. Csisz\'{a}r: {\em Studia\ Sci.\ Math.\ Hungar.\/} {\bf 2}, 299--318 (1967).

\bibitem{petz86}
D. Petz: {\em Rep. Math. Phys.\/} {\bf 21}, 57--65 (1986).

\bibitem{audn13}
K. M. R. Audenaert: {\em Quantum. Inf. Comput.\/} {\bf 14}, 0031--0038 (2013).

\bibitem{cbr13}
N. Ciganovi\'{c}, N. J. Beaudry, and R. Renner: {\em Smooth max-information as one-shot
generalization for mutual information\/}, arXiv:1308.5884 [math-ph] (2013).

\bibitem{CT91}
T. M. Cover and J. A. Thomas: {\em Elements of Information Theory\/}, John Wiley {\&} Sons, New York 1991.

\bibitem{EP04}
D. Erdogmus and J. C. Principe: {\it J.\ VLSI Signal Process.\/} {\bf 37}, 305--317 (2004).

\bibitem{sf06}
S. Furuichi: {\it J.\ Math.\ Phys.\/} {\bf 47}, 023302 (2006).

\bibitem{renyi61}
A. R\'{e}nyi: {\em On Measures of Entropy and Information}, in Proc. 4th Berkeley Symp. on Mathematical
Statistics and Probability, J.~Neyman ed., 547--561, University of California Press, Berkeley, CA 1961.

\bibitem{tsallis}
C. Tsallis: {\em J.\ Stat.\ Phys.\/} {\bf 52}, 479--487 (1988).

\bibitem{rastkyb}
A. E. Rastegin: {\em Kybernetika} {\bf 48}, 242--253 (2012).

\bibitem{levitin99}
L. B. Levitin: {\em Chaos, Solitons {\&} Fractals\/} {\bf 10}, 1651--1656 (1999).

\bibitem{schrader00}
R. Schrader: {\em Fortschr. Phys.} {\bf 48}, 747--762 (2000).

\bibitem{lieb75}
E. H. Lieb: {\em Bull.\ Am.\ Math.\ Soc.\/}. {\bf 81}, 1--13 (1975).

\bibitem{LR73}
E. H. Lieb and M. B. Ruskai: {\em J.\ Math.\ Phys.\/} {\bf 14}, 1938--1941 (1973).

\bibitem{nielsen}
M. A. Nielsen and I. L. Chuang: {\em Quantum Computation and Quantum Information\/}, Cambridge University Press, Cambridge 2000.

\bibitem{rren05}
R. Renner: {\em Security of quantum key distribution\/}, PhD thesis, ETH Zurich, arXiv: quant-ph/0512258 (2005).

\bibitem{krs09}
R. K\"{o}nig, R. Renner and C. Schaffner: {\em IEEE Trans.\ Inf.\ Theory\/} {\bf 55}, 4337--4347 (2009).

\bibitem{mdsft13}
M. M\"{u}ller-Lennert, F. Dupuis, O. Szehr, S. Fehr and M. Tomamichel:
{\em J.\ Math.\ Phys.\/} {\bf 54}, 122203 (2013).

\bibitem{giro13}
N. Gigena and R. Rossignoli: {\em J.\ Phys.\ A: Math.\ Theor.\/} {\bf 47}, 015302 (2014).

\bibitem{ics08}
I. Csisz\'{a}r: {\em Entropy} {\bf 10}, 261--273 (2008).

\bibitem{drag03}
S. S. Dragomir: {\em Appl.\ Math.\/} {\bf 48}, 205--223 (2003).

\bibitem{wwy13}
M. M. Wilde, A. Winter, and D. Yang: {\em Strong converse for the classical capacity of
entanglement-breaking and Hadamard channels\/}, arXiv:1306.1586 [quant-ph] (2013).

\bibitem{adat13}
K. M. R. Audenaert and N. Datta: {\em $\alpha$-$z$-relative R\'{e}nyi entropies\/}, arXiv:1310.7178 [quant-ph] (2013).

\bibitem{ww13}
M. M. Wilde and A. Winter: {\em Strong converse for the classical
capacity of the pure-loss bosonic channel\/}, arXiv:1308.6732
[quant-ph] (2013). 

\bibitem{fl13}
R. L. Frank and E. H. Lieb: {\em J.\ Math.\ Phys.\/} {\bf 54}, 122201 (2013). 

\bibitem{sb13}
S. Beigi: {\em J.\ Math.\ Phys.\/} {\bf 54}, 122202 (2013).

\bibitem{mosog13}
M. Mosonyi and T. Ogawa: {\em Quantum hypothesis testing and the
operational interpretation of the quantum R\'{e}nyi relative
entropies\/}, arXiv:1309.3228 [quant-ph] (2013).

\bibitem{rfz10}
W. Roga, M. Fannes and K. \.{Z}yczkowski: {\em Phys.\ Rev.\ Lett.\/} {\bf 105}, 040505 (2010).

\bibitem{rzf11}
W. Roga, M. Fannes, and K. \.{Z}yczkowski: {\em Int.\ J.\ Quantum Inf.\/} {\bf 9}, 1031--1046 (2011).

\bibitem{rast11a}
A. E. Rastegin: {\em J.\ Phys.\ A: Math.\ Theor.\/} {\bf 45}, 045302 (2012).

\bibitem{rastcejp}
A. E. Rastegin: {\em Cent.\ Eur.\ J.\ Phys.\/} {\bf 11}, 69--77 (2013).

\bibitem{rastjst12}
A. E. Rastegin: {\em J.\ Stat.\ Phys.\/} {\bf 148}, 1040--1053 (2012).

\bibitem{alf04}
R. Alicki and M. Fannes: {\em J.\ Phys.\ A: Math.\ Gen.\/} {\bf 37}, L55--L57 (2004).

\bibitem{marco12}
M. Tomamichel: {\em A framework for non-asymptotic quantum information theory\/}, PhD thesis, ETH Zurich, arXiv:1203.2142 [quant-ph] (2012).

\bibitem{bhatia97}
R. Bhatia: {\em Matrix Analysis\/}, Springer, New York 1997.

\bibitem{borland}
L. Borland, A. R. Plastino, and C. Tsallis: {\em J.\ Math.\ Phys.\/} {\bf 39}, 6490--6501 (1998).

\bibitem{ics66}
I. Csisz\'{a}r: {\em Studia Sci.\ Math.\ Hungar.\/} {\bf 1}, 227--230 (1966).

\bibitem{ggil}
G. Gilardoni: {\em IEEE Trans.\ Inf.\ Theory\/} {\bf 56}, 5377--5386 (2010).

\bibitem{rastmpag}
A. E. Rastegin: {\em Math.\ Phys.\ Anal.\ Geom.\/} {\bf 16}, 213--228 (2013).

\bibitem{akr13}
A. K. Rajagopal, Sudha, A. S. Nayak, and A. R. Usha Devi: {\em Phys. Rev. A} {\bf 89}, 012331 (2014).

\bibitem{ar01}
S. Abe and A. K. Rajagopal: {\em Physica A\/} {\bf 289}, 157--164 (2001).

\bibitem{raggio}
G. A. Raggio: {\em J.\ Math.\ Phys.\/} {\bf 36}, 4785--4791 (1995).

\bibitem{bhatia07}
R. Bhatia: {\em Positive Definite Matrices\/}, Princeton University Press, Princeton 2007.

\bibitem{rockaf}
R. T. Rockafellar: {\em Convex Analysis\/}, Princeton University Press, Princeton 1970.


\end{thebibliography}
\end{document}